\documentclass[english,aps,prd,showkeys,print,amsfonts,amsmath,amssymb,superscriptaddress]{revtex4-1}
\usepackage[T1]{fontenc}
\usepackage[latin9]{inputenc}
\usepackage{babel}
\usepackage{verbatim}
\usepackage{bm}
\usepackage{amsmath}
\usepackage{textcomp}
\usepackage{ragged2e}
\usepackage{stackrel}
\usepackage{graphicx}
\usepackage[unicode=true,
 bookmarks=false,
 breaklinks=false,pdfborder={0 0 1},backref=section,colorlinks=false]
 {hyperref}

\makeatletter

\providecommand{\tabularnewline}{\\}

\makeatother

\begin{document}

%\linenumbers

\title[Article Title]{Global Gravity Field Model from Taiji-1 Observations}

\author{Liming Wu}
\affiliation{Department of Geophysics,
Chang'an University, Middle-section of Nan'er Huan Road, Xi'an 710064, China.}
\affiliation{Institute of Mechanics, Chinese Academy of Sciences, No.15 Beisihuanxi Road, Beijing 100190,
China.}
\author{Peng Xu}\thanks{Corresponding author.}
\email{xupeng@imech.ac.cn}
\affiliation{Institute of Mechanics, Chinese Academy of Sciences, No.15 Beisihuanxi Road, Beijing 100190,
China.}
\affiliation{Lanzhou Center of Theoretical Physics, Lanzhou University, No. 222 South Tianshui Road, Lanzhou
730000, China.}
\affiliation{Hangzhou Institute for Advanced Study, University of Chinese Academy
of Sciences, 84 Church Street SE, Hangzhou 310024, China.}
\author{Shuhong Zhao}
\affiliation{Department of Geophysics,
Chang'an University, Middle-section of Nan'er Huan Road, Xi'an 710064, China.}
\author{Li-E Qiang}
\affiliation{National Space Science Center, Chinese Academy of Sciences, NO.1 Nanertiao Zhongguancun,
Beijing 100190, China.}
\author{Ziren Luo}
\affiliation{Institute of Mechanics, Chinese Academy of Sciences, Beijing, 100190,
China.}
\author{Yueliang Wu}
\affiliation{University of Chinese Academy of Sciences, No.19(A) Yuquan Road, Beijing 100049, China.}


\begin{abstract}
Taiji-1 is the first technology demonstration satellite of the Taiji program of China's space-borne gravitational wave antenna. After the demonstration of the key individual technologies, Taiji-1 continues
    collecting the data of the precision orbit determinations, satellite
    attitudes, and non-conservative forces exerted on the S/C. Therefore,
    during its free-fall, Taiji-1 can be viewed as operating in the high-low
    satellite-to-satellite tracking mode of a gravity recovery mission.
    In this work, we have selected and analyzed the one month data from Taiji-1's observations, and developed the techniques to resolve the long term
    interruptions and disturbances in the data due to the
    scheduled technology demonstration experiments.
    The first global gravity model \texttt{TJGM-r1911}, that independently derived from China's own satellite mission, is successfully built from Taiji-1's observations.
    Compared with gravity models from CHAMP and other satellite gravity missions,
    the accuracy discrepancies exist, which is mainly caused by the data
    discontinuity problem. As the extended free-falling phase been approved, Taiji-1 could serve as a gravity recovery mission for China since 2022 and it will provide us the
    independent measurement of both the static and
    the monthly time-variable global gravity field.
\end{abstract}

\keywords{Taiji program, Gravity field recovery, Satellite gravity, Gravitational wave detection}



\maketitle

\section{Introduction\label{sec:Introduction}}

In 2000, Chinese Academy of Sciences (CAS) established China's first working group of the space borne gravitational wave observatories that led by Academician Wen-Rui Hu.
China had then started her own journey to gravitational wave detections in space and joined the international corporations led by the LISA team \cite{danzmann_lisa_2011}.
Motivated mainly by the concept of ALIA mission \cite{bender_additional_2004}, China's first concept of space borne gravitational wave antennas was proposed in 2011 \cite{gong_scientific_2011}, and afterward, a more conservative design was made  \cite{gong_descope_2015,xue-fei_laser_2015}.
In 2016, the breaking news of the first detection of gravitational wave \cite{the_ligo_scientific_collaboration_advanced_2015,abbott_observation_2016,abbott_observing_2016,abbott_properties_2016} was announced by the Adv-LIGO team, which was soon recognized as one of the most significant achievements in this new century of general relativity.
The subsequent gravitational wave detections by the LIGO-VIRGO collaboration had then raised the curtain of the new era of gravitational wave astronomy and astrophysics, and also boosted the progress of the space missions.
Along with such breakthroughs, and also being encouraged by the successful experiments of the LISA Pathfinder mission \cite{armano_sub-femto-_2016,armano_beyond_2018,armano_lisa_2019,armano_lisa_pathfinder_2019,anderson_experimental_2018}, the Taiji program in space was released by CAS in 2016 \cite{cyranoski_chinese_2016,hu_taiji_2017}
which outlined its 3-step R\&D roadmap for China's space gravitational wave antenna in the future \cite{luo_taiji_2020,luo_brief_2020}.
The ultimate goal of this program is the Taiji mission,
a heliocentric LISA-like mission that expected to be launched in the early 2030s.
Consisting of three space-crafts (S/C), the Taiji mission will be an almost
equilateral triangular constellation with the arm-length about $3\times10^{6}$
km and the sensitive band ranging from 0.1 mHz to 1 Hz.
The scientific objectives will include such gravitational wave sources as coalescing supermassive black hole binaries, extreme mass ratio inspirals, stochastic gravitational wave backgrounds, etc. \cite{hu_taiji_2017,ruan_taiji_2020,luo_taiji_2020,luo_brief_2020,the_taiji_scientific_collaboration_taiji_2021,ruan_lisataiji_2020,ruan_lisa-taiji_2021,wang_hubble_2021,wang_alternative_2021}.

With the phase A study started in May 2018 and the ground-based tests of related technologies \cite{Liuheshan2018,Liuheshan2018a,Liyuqiong2018,Liyupeng2019,Liyuqiong2020,Liuhang2021}, the Taiji-1 mission, as the first technology demonstration satellite of the Taiji program, was approved by CAS in August 2018. The detailed design was finished in October 2018 and all the unit flight models were delivered before APR 2019.
The assembly integrations and tests were then conducted from APR 2019, and in August 2019 Taiji-1 was launched.
The most important individual technologies of China's space
gravitational wave antenna were verified in space, including gravitational reference
sensors (GRS), high precision laser interferometers, drag-free
control system, $\mu$-N thrusters, and the ultra-stable and clean platform.
The successful operation of Taiji-1 had demonstrated and confirmed the designed performances of the scientific payloads and the satellite platform \cite{the_taiji_scientific_collaboration_chinas_2021,the_taiji_scientific_collaboration_taiji_2021}. In 2022, the Taiji-1 satellite has entered into its final extended phase,
and the challenging experiment on global gravity recoveries will be carried out.

The orbital dynamics of
Taiji-1 is determined by the forces acting on the S/C, which, due
to their physical origins, can be divided into three classes: the
gravitational forces including the almost centripetal force from the
Earth and the perturbations from the Sun, the Moon and other celestial
bodies, the non-conservative forces from the space environments such
as air drags, Solar radiation pressure Earth albedo, and also
the disturbances from thruster events.
During its science operation, Taiji-1 has continuously collected the precision orbit determination (POD) data based on the Global Navigation
Satellite Systems (GPS and BeiDou), the satellite attitude data from star sensors,
and also the precision measurements by the GRS of the non-conservative
forces exerted on the S/C.
Therefore, given such observation data, the
detailed information of the Earth gravitational field could be inferred
based on the satellite dynamic model.

\begin{figure}[ht]
\centering \includegraphics[width=0.48\textwidth]{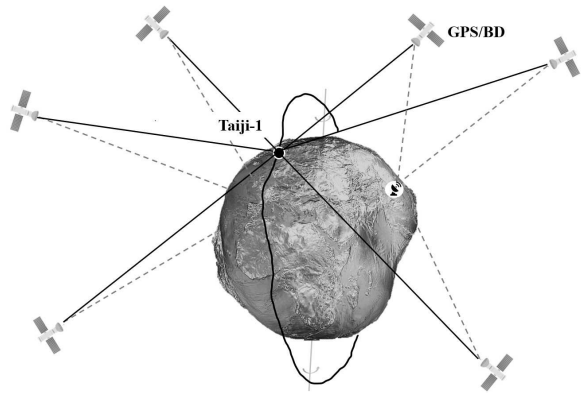} \caption{The illustration of the high-low satellite-to-satellite tracking mode
of Taiji-1.}
\label{fig:HLSST}
\end{figure}

From this point of view, apart from the disruptions by the technology
demonstration experiments such as the thruster performance tests,
the drag-free control tests, the satellite maneuvers and so on, the
free-falling mode of the Taiji-1 satellite can be viewed as the high-low
Satellite-to-Satellite Tracking (hl-SST) mode of the Earth gravity
recovery mission \cite{moore_champ_2003,naeimi_global_2017,reigber_earth_2005-1}.
See Fig. \ref{fig:HLSST} for illustration.

Taiji-1 could then serve as a gravity recovery satellite for
China, which could provide us the independent measurements of both the static global gravity field over about one year and the monthly averaged time-variable gravity field of long wavelengths.

The gravity model obtained from Taiji-1, as the tentative data products,
could then fuse with the up-coming geopotential measurements from
the official Low-Low SST gravity recovery mission of China's satellite
gravity program. Further, such precedent data products could also
provide the opportunity to make valuable cross-validations between
the two missions. Based on such considerations, the Taiji team had approved the extended free-falling phase of Taiji-1 in the end of 2021, and the satellite is to follow its geodesic orbit around Earth with minimal disruptions and disturbances only from events as attitude adjustments, etc.

In this work, we introduce the first Earth gravity field product
obtained from Taiji-1's observations during its science operation.
The conventional energy integral method employed in gravity recoveries
for hl-SST missions is outline in Sec. \ref{sec:Energy-integral-method}.
Since the free-falling motion of the S/C in Earth's gravitational
field and the continual observations of the orbit and non-conservative
forces are crucial to the global gravity recoveries, the data set
is carefully selected to avoid the disruptions and gaps as much as
possible. The measurement data from Taiji-1 and the ancillary models
required are described in Sec. \ref{sec:Required-data-set}. While
the interruptions and the anomalies still exist in the chosen data,
the software package \texttt{TJGrav} is developed to resolve such
problems and to process the data. The data fusion techniques are developed
to synthesize the measurement data with certain models that are calibrated
carefully by the measurements. The data processing procedure can be
found in Sec. \ref{sec:Data-processing}.The first monthly
Earth gravity model \texttt{TJGM-r1911} independently derived from
China's own satellite mission Taiji-1 is discussed in Sec. \ref{sec:Results}.
Conclusions and an outlook are provided in Sec. \ref{sec:Conclusions}.

\section{Energy integral method\label{sec:Energy-integral-method}}

Proposed by O'keefe in \cite{okeefe_application_1957} firstly and
investigated in detail in \cite{hotine_first_1969,wolff_direct_1969,jekeli_determination_1999,visser_energy_2003},
the energy integral or the energy balance approach, has been developed
as a full-fledged and widely used method in the global gravity inversions
for hl-SST satellite gravity missions. It is based on the theoretical
prediction that ideally the Jacobi's integral of a S/C motion is conserved
along its orbits \cite{jacobi_uber_1836}, which can be viewed as
an equivalent to the conservation of total mechanical energy. Therefore,
the balance between the kinetic energy and the gravitational energy
of the S/C can be used to derive the detailed information of the gravity
field, given the measured data of the S/C position and velocity.

The key to the approach in realistic applications is to account for
the energy dissipation caused by the non-conservative surface forces
acting on the S/C accurately. Generally, low Earth orbits are adopted
for the satellite gravity missions, and the main contributions to
the total non-conservative force include the air drags, the solar
radiation pressures and also the Earth albedo pressure. One can use
force models to reduce the errors caused by such energy loss, while
the precise and real time measurements of such perturbation forces
will greatly improve the fitting accuracy of the gravity field. Please
see \cite{visser_energy_2003,gerlach_champ_2003,gerlach_champonly_2003,moore_champ_2003,reigber_gravity_2005,jaggi_gps-only_2011}
for the detailed discussions of the energy integral method and its
applications to the CHAMP, GRACE and GOCE missions.

Here, we outline the theoretical principle of the energy integral
method and give the necessary definitions used in this work. The functional
model chosen here to estimate the geopotential $V$ is the following
spherical harmonic series truncated at maximum degree $N$ and defined
in the Earth centered spherical coordinates system $\{r,\theta,\lambda\}$
(radius, geocentric latitude, longitude),
\begin{equation}
V\left(r,\theta,\lambda\right)=\frac{GM}{r}\left[1+\sum_{n=2}^{N}\sum_{m=-n}^{n}A_{nm}\left(\frac{R}{r}\right)^{n}Y_{nm}\left(\theta,\lambda\right)\right],\label{eq1}
\end{equation}
where $G$ is the gravitational constant, $M$ the total mass of the
Earth, and $R$ the mean equatorial radius. $Y_{nm}$ denotes the
surface spherical harmonics of degree $n$ and order $m$
\begin{equation}
Y_{nm}(\vartheta,\lambda)=\bar{P}_{n\lvert m\rvert}(\sin\theta)\begin{cases}
\cos m\lambda & m\ge0\\
\sin\lvert m\rvert\lambda & m<0,
\end{cases}\label{eq2}
\end{equation}
where $\bar{P}_{nm}$ is the fully normalized associated Legendre
function of degree $n$ and order $m$, and $A_{nm}=\{C_{nm},S_{nm}\}$,
with $C_{nm}$ for $m\geq0$ and $S_{nm}$ for $m<0$, are the
unknown spherical harmonic coefficients to be determined.

The Earth gravity field, in terms of the geopotential coefficients
$A_{mn}$, could be obtained from the least-square solutions of the
observation equations that link the orbit position and velocity solutions
to the gravity field unknowns. The energy integral is used to derive
the observation equations, which is defined along the S/C orbit as
\begin{equation}
V=V_{0}+\int_{t_{0}}^{t}\boldsymbol{g}\cdot\boldsymbol{\dot{r}}d\tau,\label{eq4}
\end{equation}
where $V$ and $V_{0}$ are the Earth geopotentials at the S/C orbit
positions of time $t$ and $t_{0}$ respectively, $\boldsymbol{r}$
and $\dot{\boldsymbol{r}}$ the position and velocity vector of the
S/C in the Earth-centered and Earth-fixed reference frame.
The vector $\boldsymbol{g}$ denotes the acceleration caused by static
geopotentials of the Earth
\begin{equation}
\boldsymbol{g}=\boldsymbol{\ddot{r}}+2\boldsymbol{\omega}\times\boldsymbol{\dot{r}}+\boldsymbol{\omega}\times\left(\boldsymbol{\omega}\times\boldsymbol{r}\right)-\boldsymbol{a_{NC}-a_{G},}\label{eq5}
\end{equation}
where $\boldsymbol{\omega}$ is the angular velocity of the Earth
rotation. $ \boldsymbol{a_{NC}}$ includes all the non-gravitational accelerations
exerted by the satellite, and\textbf{ $\boldsymbol{a}_{\boldsymbol{G}}$}
counts for all the perturbations from other time-varying gravitational
sources such as tides, mass transfers in atmosphere and ocean,
and 3rd-body effects from the Sun or the Moon. To resolve the long
wavelength or low-degree gravity field, we split the Earth geopotentials
into three parts $U$, $V_{l}$ and $V_{h}$, where $U$ represents
the monopole potential, $V_{l}$ is the potential consisting of low-degree
harmonics and $V_{h}$ the one from high-degree harmonics. Thus, Eq.\ref{eq4}
becomes
\begin{align}
H+V_{l}\left(\boldsymbol{r}\right)= & \frac{1}{2}\mathbf{\dot{r}}\cdot\mathbf{\dot{r}}-\frac{1}{2}\left(\boldsymbol{\omega}\times\boldsymbol{r}\right)\cdot\left(\boldsymbol{\omega}\times\boldsymbol{r}\right)\nonumber \\
 & -\int_{t_{0}}^{t}\boldsymbol{a_{NC}}\cdot\mathbf{\dot{r}}\mathrm{d}\tau-\int_{t_{0}}^{t}\boldsymbol{a_{G}}\cdot\mathbf{\dot{r}}\mathrm{d}\tau-U-V_{h}\label{eq3}
\end{align}
The unknown integration constant $H$ and the coefficients $\{C_{nm},\ S_{nm}\}$
of the low-degree components on the left-hand side of Eq. \ref{eq3}
are to be resolved with Taiji-1's observations and the modeled
ancillary data substituted into the right-hand side of the equation.

\section{Data sets and ancillary models \label{sec:Required-data-set}}

\subsection{Data sets}

The Taiji-1 satellite was launched to a circular Sun-synchronous dawn/dusk
orbit with the altitude about $600\ km$ and inclination angle $97.67^{o}$.
The orbit has a stable Sun-facing angle, which can provide a constant
power supply for the battery and also the stable temperature gradient
for the platform. The satellite is about 180 kg, and along such orbit, the
solar radiation pressure contributes to the non-conservative forces
dominantly, and the air drags along the orbit turns out to be small.
One of the key payloads of Taiji-1 is the GRS installed at the mass
center of the S/C. Except for the drag-free control experiments during
the science operation, where the satellite is controlled to trace
the motions of the test mass inside the GRS along the radial
direction, the GRS are set to work in the accelerometer mode in most
cases. The electrostatic actuation forces keeping the test mass to follow
the motions of the S/C will then provide the precise measurements
of the non-gravitational forces exerted by the S/C.

The data products needed for the Earth gravity recoveries include
the POD data from both the GPS and BeiDou (BD) systems, the GRS data of
the non-gravitational forces measured in the satellite frame
and the S/C attitude data to transform the physical measurements from the satellite frame
to the inertial reference frame. In this work, our first Taiji-1 gravity
model is based on the data from 01-11-2019 to 31-11-2019. As mentioned,
during this month fewer experiments related to the GRS, the thrusters,
and the satellite-maneuvers were performed. The S/C maintained steadily
in the Earth pointing attitude and the POD data is in good quality. The one-month data length was chosen since any extension could hardly improve
the final fitting accuracy due to the frequent disruptions and gaps
in the data from the technology demonstration experiments. Further,
since the monthly measurements of the time-variable gravity field
from Taiji-1's extended free-falling phase is under preparation, this
first monthly gravity model could serve as a reference.

\begin{widetext}

\begin{figure}[ht]
\centering
\includegraphics[width=0.9\textwidth]{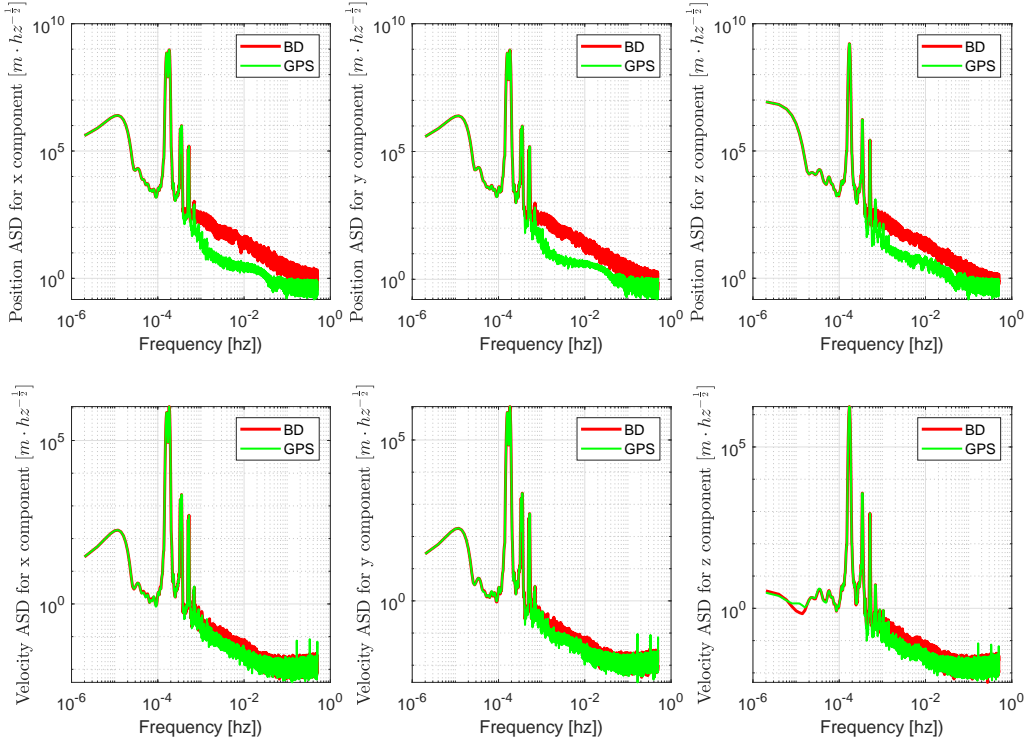}
\caption{ASDs of Taiji-1's POD data from both GPS and BD in the Earth-centered and Earth-fixed reference frame.\label{fig:POD ASD}}
\end{figure}

\end{widetext}

\begin{figure}[ht]
\centering \includegraphics[width=0.48\textwidth]{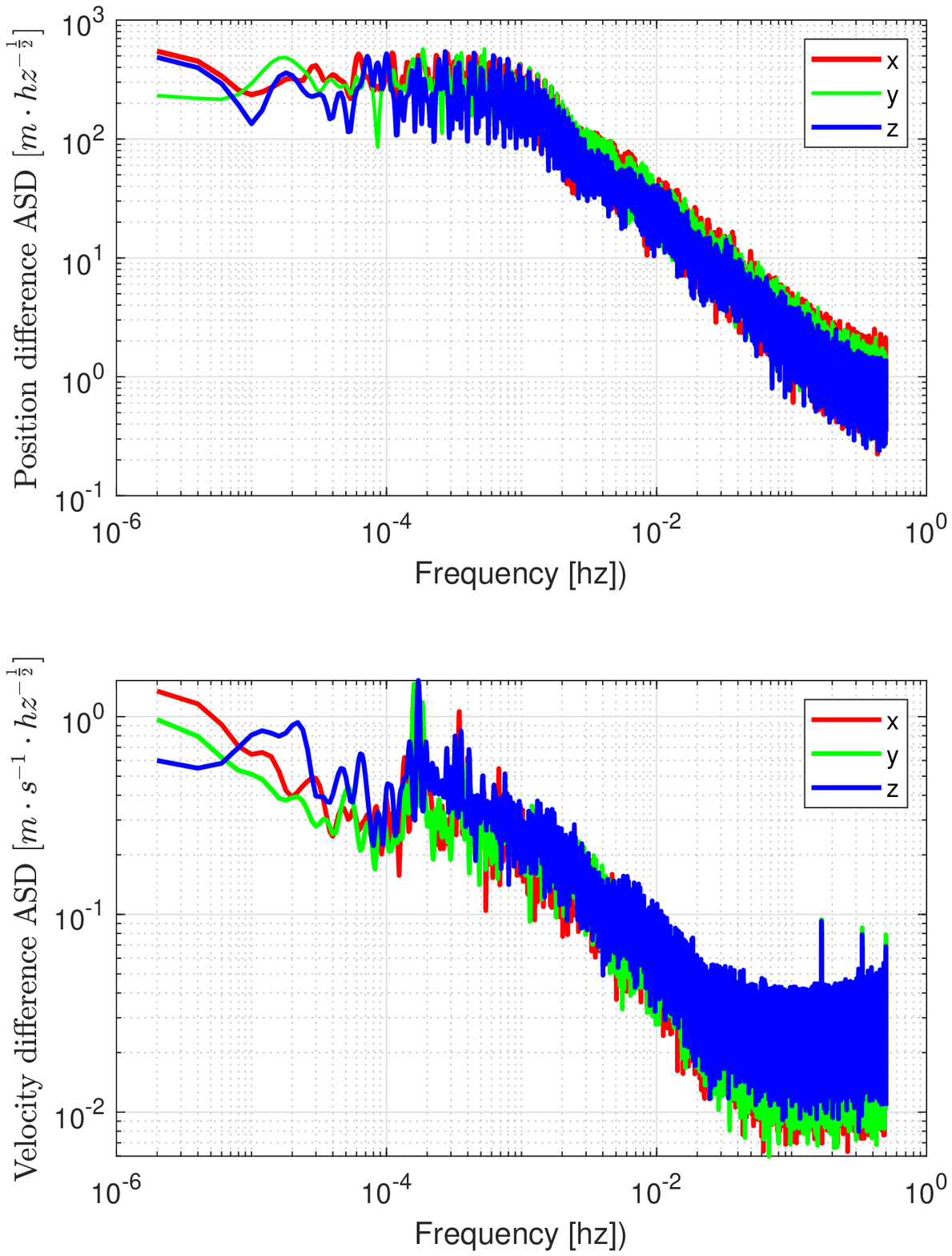}
\caption{ASDs of the POD data differences between BD and GPS in the Earth-centered and Earth-fixed reference frame.\label{fig:POD diff}}
\end{figure}

The orbit precision of Taiji-1 is determined by the
Global Navigation Satellite Systems, including both the GPS and the BD system. The POD data from GPS and BD are both defined in the Earth-centered and Earth-fixed reference frame with the sampling rate of 1 Hz.
The amplitude spectrum density (ASD) of the POD data including
the positions and the velocities of Taiji-1 are shown in Fig. \ref{fig:POD ASD}. In most cases, the Taiji-1 satellite can
be more often tracked by GPS satellites than by the BD satellites.
Fig. \ref{fig:POD diff} shows the difference between the measurements
from BD and GPS.

\begin{figure}[ht]
\centering \includegraphics[width=0.48\textwidth]{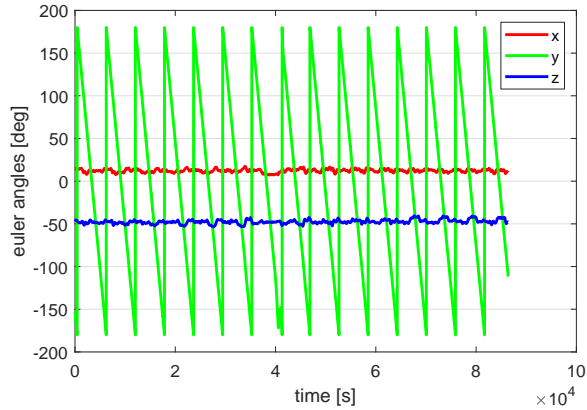}
\caption{Time series of the S/C attitude data (Euler angles) on 01-11-2019.
\label{fig:Eular}}
\end{figure}

The S/C attitude is measured by star sensors, and
the detailed information of how to rotate from the satellite frame
to the inertial frame is provided by the Euler angle $\{\theta,\ \phi,\ \psi\}$
data product. In Fig. \ref{fig:Eular} the time series of the S/C
attitude on 01-11-2019 are shown. The rotation matrices for each axis are given by
\begin{eqnarray}
\mathbf{R}_{x}(\theta) & = & \left(\begin{matrix}1 & 0 & 0\\
0 & \cos\theta & \sin\theta\\
0 & -\sin\theta & \cos\theta
\end{matrix}\right)
\end{eqnarray}
\begin{eqnarray}
\mathbf{R}_{y}(\phi) & = & \left(\begin{matrix}\cos\phi & 0 & -\sin\phi\\
0 & 1 & 0\\
\sin\phi & 0 & \cos\phi
\end{matrix}\right)
\end{eqnarray}
\begin{eqnarray}
\mathbf{R}_{z}(\psi) & = & \left(\begin{matrix}\cos\psi & \sin\psi & 0\\
-\sin\psi & \cos\psi & 0\\
0 & 0 & 1
\end{matrix}\right)
\end{eqnarray}
The total transformation matrix from the inertial reference frame
to the satellite frame reads
\begin{equation}
\boldsymbol{R}=\boldsymbol{R}_{x}\boldsymbol{R}_{y}\boldsymbol{R}_{z}\label{eq7}
\end{equation}

The non-conservative forces are
measured by the GRS in the accelerometer mode.
The sensor unit of the electrostatic GRS on-board Taiji-1 contains two parts: the mechanical assembly and the front-end electronics unit.
The mechanical assembly consists of a $72\ g$ parallel hexahedral titanium alloy
test mass and an electrode cage that encloses the test mass.
The test mass, as the inertial reference, is suspended electrostatically inside the cage.
When non-gravitational forces acting on the S/C cause the relative motions between the test mass and the cage, the capacitance between the test mass and the electrodes changes slightly and induces signals that could be picked up by the front-end electronics unit.
Based on the position sensor data, the test mass maintained its nominal position inside the cage by applying low frequency actuation voltages through the electrodes.
The voltages sampled at 100 Hz are then transformed, with
the calibrated bias and the scale factors, into the non-conservative
accelerations exerted by the S/C in the satellite frame.

\begin{figure}[ht]
\centering \includegraphics[width=0.48\textwidth]{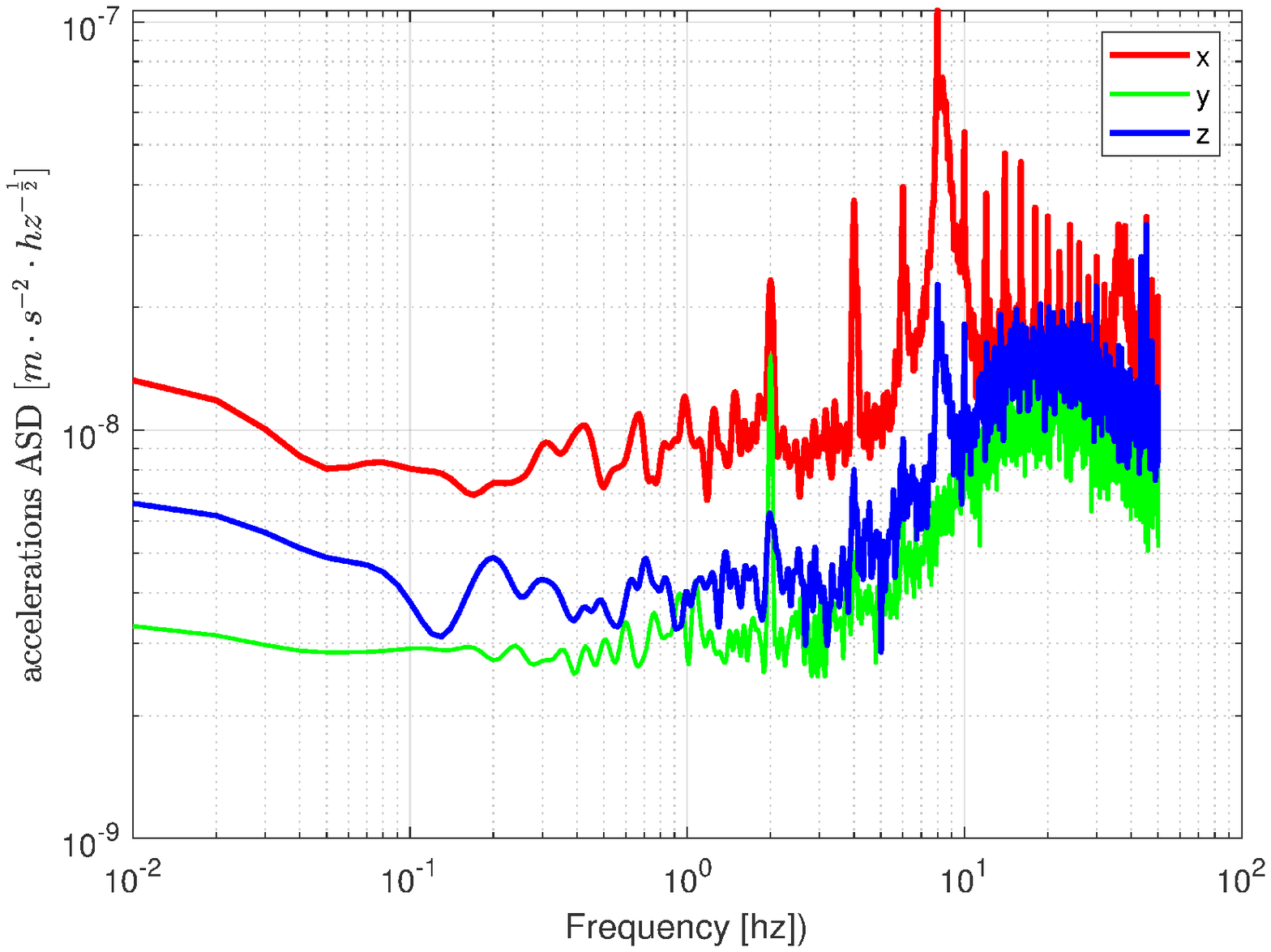}
\caption{ASDs of the measured non-gravitational accelerations in the satellite frame.\label{fig:ASDACC}}
\end{figure}

The sensitivities of the three GRS axes are different. The x-axis is pointing towards the Earth with low sensitivity, while the y-axis and z-axis
are along the flight direction and the orbital normal direction respectively with high sensitivities. According to the in-orbit performance tests
\citep{the_taiji_scientific_collaboration_chinas_2021,min_performance_2021,peng_system_2021},
the resolutions of the high sensitive axes reach the level of $\text{1\ensuremath{0^{-10}}m/\ensuremath{s^{2}}/H\ensuremath{z^{1/2}}}$
in the frequency band from 0.01 Hz to 1 Hz and $10^{-9}m/s^{2}/Hz^{1/2}$
in the low frequency band from 0.1 mHz to 0.01 Hz. It will provide
us the faithful measurements of the non-gravitational accelerations
along Taiji-1's orbit, according to the known force models. See Fig.
\ref{fig:SimACC} in the next section. Fig. \ref{fig:ASDACC} shows
the ASDs of the measured non-gravitational accelerations in the satellite frame.

\subsection{Ancillary models}

% \begin{widetext}

\begin{table}
\centering
\caption{Models for the perturbation forces\label{tab:Models-for-the}}
\begin{tabular}{|c|c|c|}
\hline
{\scriptsize Perturbation}  & {\scriptsize Model}  & {\scriptsize Maximum degree } \tabularnewline
\hline
\hline
{\scriptsize Earth's gravity field}  &{\scriptsize EGM2008}  &{\scriptsize degree/order 100} \tabularnewline
\hline
{\scriptsize Dealiasing}  & {\scriptsize GRACE AOD1B RL06 } &{\scriptsize degree/order 50} \tabularnewline
\hline
{\scriptsize Ephemerides }  &{\scriptsize JPL DE421}  & \tabularnewline
{\scriptsize of Sun and Moon} & & \tabularnewline
\hline
{\scriptsize Solid Earth tides}  &{\scriptsize IERS2010}  &{\scriptsize degree/order 4} \tabularnewline
\hline
{\scriptsize Relativistic corrections}  &{\scriptsize IERS2010}  & \tabularnewline
\hline
{\scriptsize Atmospheric density}  &{\scriptsize NRLMSISE-00}  & \tabularnewline
\hline
{\scriptsize Earth Albedo}  &{\scriptsize CERES}  & \tabularnewline
\hline
\end{tabular}
\end{table}

% \end{widetext}
To determine the long wavelength and low-degree gravity field model,
the interference from the high-degree geopotentials, the tidal signals
and the 3rd-body perturbations needs to be suppressed or removed.
In this work, the EGM2008 model \citep{pavlis_development_2012} is
employed as the prior gravity model,
and also one of the reference models to assess and
validate our Taiji-1 gravity model. Solid-Earth tides and the relativistic
corrections are also considered according to the standard IERS2010
\citep{petit_iers_2011}. The 3rd-body perturbations from the Sun
and the Moon are included, and the required position and the velocity
vectors of the Sun and the Moon are obtained from JPL's ephemerids
DE421 \citep{folkner_planetary_2009}. The non-tidal atmospheric and
oceanic mass variations are evaluated based on the atmosphere and
ocean dealiasing product AOD1B RL06 \citep{dobslaw_new_2017}.

As mentioned, though fewer in the selected data, disruptions and gaps still exist, especially in the measurements of the GRS system.
Therefore, the simulations
of the non-gravitational forces calibrated and adjusted by the measurements
along the orbits will be employed as the complementary data when the
disturbances or the long term interruptions happen. In this work,
the atmospheric density model NRLMSIS-00 \citep{picone_nrlmsise-00_2002}
and Earth albedo model CERES \citep{wielicki_clouds_1998,wielicki_changes_2005}
are used. The full list of all the models used can be found in Table \ref{tab:Models-for-the}.

\section{Data processing\label{sec:Data-processing}}

\begin{figure}[ht]
\centering \includegraphics[width=0.58\textwidth]{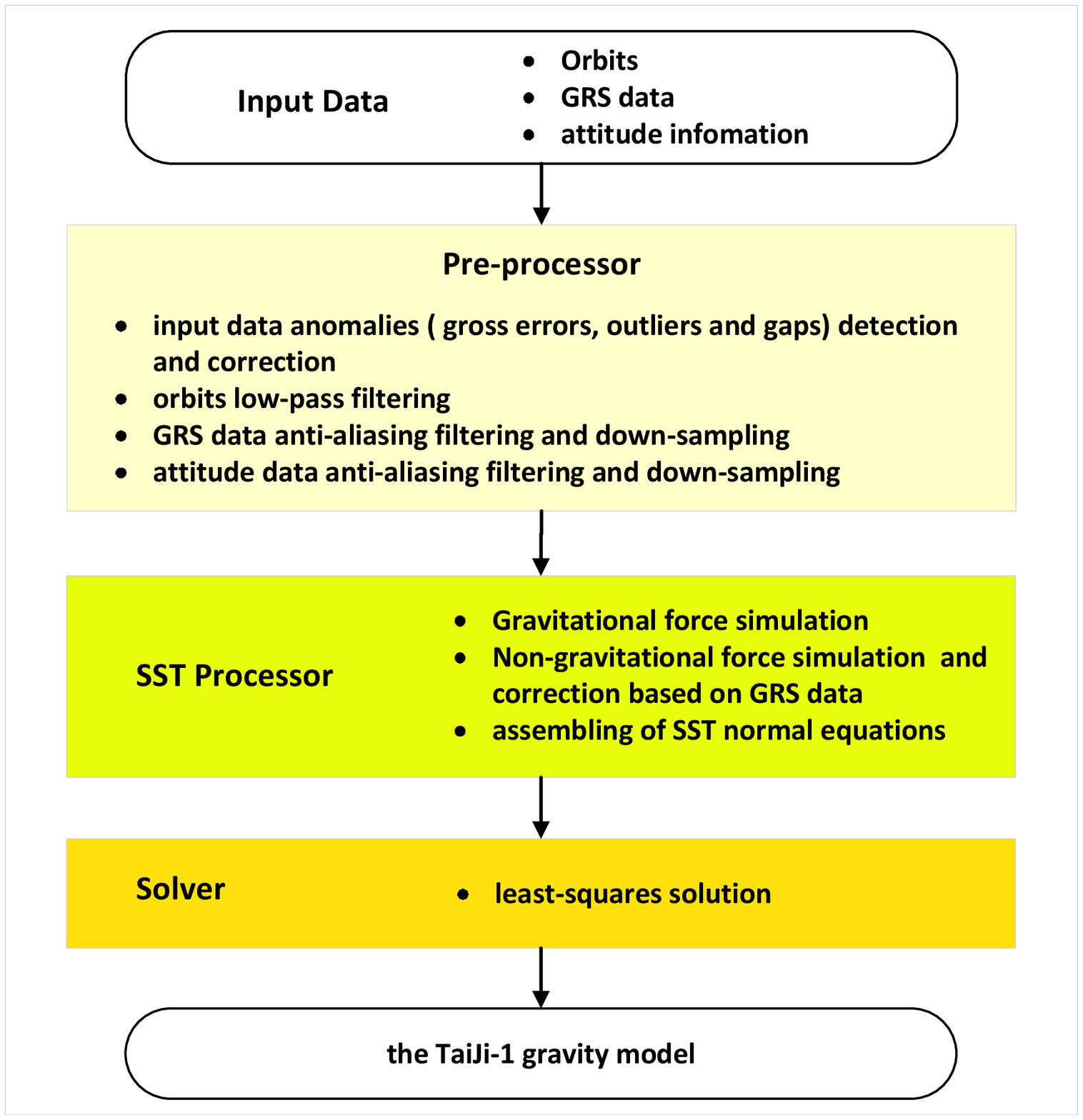}
\caption{Software architecture of \texttt{TJGrav} and the flow chart of the
data processing procedure. \label{fig:Software-architecture}}
\end{figure}

The software package \texttt{TJGrav} developed in this work for Taiji-1's
gravity field modeling contains three main sub-modules: the data
\texttt{Pre-Processor}, the \texttt{SST-Processor}, and the \texttt{Solver}.
Please see Fig. \ref{fig:Software-architecture} for the architecture
of \texttt{TJGrav} and the flow chart of the data processing procedure.

Common manipulations including the data quality check, the identifications of gaps and gross errors, adding quality flags, the anomalies removing and replaced with interpolations, data smoothing, anti-aliasing and down-samplings are carried out by the \texttt{Pre-processor.}

Take the key POD data as an example. The quality of the data are generally
good, but there are still gaps and outliers. The search for the data anomalies
is executed in the first place. To make sure the continuity of the
POD data in the observation equation \ref{eq3}, the outliers and
the gaps are replaced or filled up with values evaluated by means
of the Fourier least-square-fitting method. The replaced data will
not be used to establish the normal equations, but only to calculate
the integration in the observation equation \ref{eq3}. In the high
frequency band of the POD data, the signal-to-noise ratio of the orbital
perturbations from geopotentials decreases. Therefore, to suppress
the high frequency noises, a low-pass filter at 0.005 Hz is imposed.
After this, the POD data is down-sampled from 1 Hz to 0.2 Hz. The
similar processing is applied to the original GRS data and S/C attitude
data, and after a certain smoothing and anti-aliasing filtering they
are also down-sampled to 0.2 Hz before use.

In the \texttt{SST-Processor}, we gather all the needed ancillary
data, which are either obtained from the numerical simulations based
on well-tested models or official data products released. Afterwards,
the normal equation is established and ready to export to the \texttt{Solver}.

Required gravitational perturbations from other sources are modeled
and calculated in the Earth-centered and Earth-fixed reference frame. As mentioned in the previous sections,
the required data include perturbations from the 3rd-body (the Sun
and the Moon), solid-Earth tides, AOD, and also the relativistic corrections.
In the Earth-centered and Earth-fixed reference frame, the perturbation forces due to the 3rd-body is modeled as
\begin{equation}
\boldsymbol{a}_{\text{sun }}=-GM_{sun}\left(\frac{\boldsymbol{l}_{\text{sun }}}{\lvert\boldsymbol{l}_{\text{sun }}\rvert^{3}}+\frac{\boldsymbol{r}_{\text{sun }}}{\lvert\boldsymbol{r}_{\text{sun }}\rvert^{3}}\right)\label{eq17}
\end{equation}
\begin{equation}
\boldsymbol{a}_{\text{moon }}=-GM_{\text{moon }}\left(\frac{\boldsymbol{l}_{\text{moon }}}{\lvert\boldsymbol{l}_{\text{moon }}\rvert^{3}}+\frac{\boldsymbol{r}_{\text{moon }}}{\lvert\boldsymbol{r}_{\text{moon }}\rvert^{3}}\right)\label{eq18}
\end{equation}
where $M_{sun}$ and $M_{moon}$ are the masses of the Sun and the
Moon, $\boldsymbol{l}_{\text{sun }}$, $\boldsymbol{l}_{\text{moon }}$ are the position vectors of the S/C relative to the Sun and the Moon, and $\boldsymbol{r}_{\text{sun }}$, $\boldsymbol{r}_{\text{moon }}$ the position vectors of the Earth relative to the Sun and the Moon.
Please see Fig. \ref{fig:gravity perturbations} for the simulated
gravitational perturbations on 01-11-2019.

\begin{widetext}

\begin{figure}[ht]
\centering \includegraphics[width=0.9\textwidth]{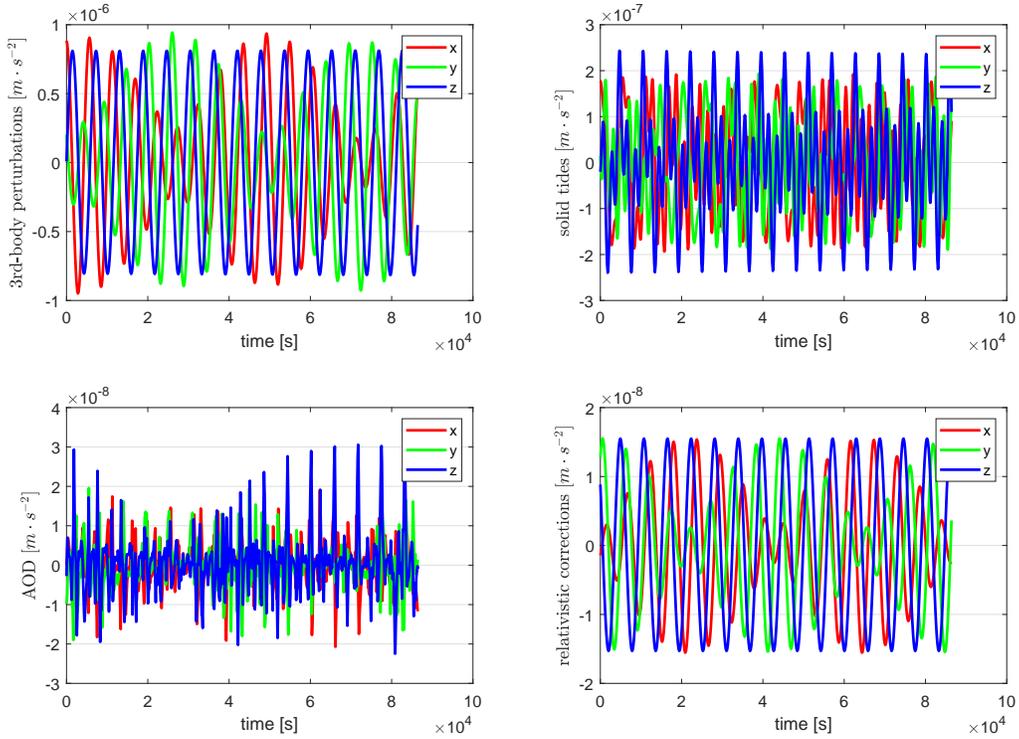}
\caption{Time series of the simulated gravitational perturbations in the earth-centered and earth-fixed reference frame
on 01-11-2019.\label{fig:gravity perturbations}}
\end{figure}

\begin{figure}[ht]
\centering \includegraphics[width=0.9\textwidth]{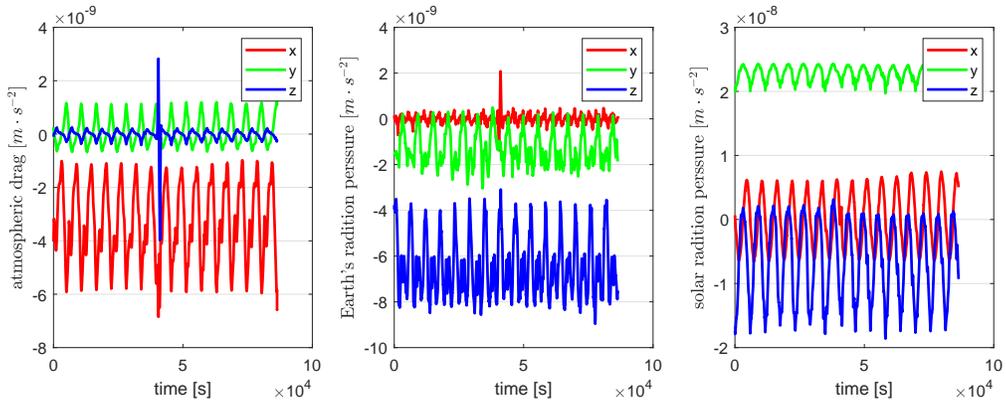}
\caption{Time series of the simulated non-gravitational forces in the earth-centered and earth-fixed reference frame
on 03-11-2019\label{fig:SimACC}}
\end{figure}

\end{widetext}

The resolution power of the GRS on-board Taiji-1 ($\sim10^{-10}m/s^{2}/Hz^{1/2}$)
makes it sensible to small disturbances from the satellite platform.
The test mass, as the inertial reference, couples to the surrounding physical
fields in complicated ways. Events like temperature variations
of the sensor unit, slight mechanical vibrations of the platforms
and so on may cause the erroneous signals in the GRS measurements. The
transient data anomalies of those kinds can be fixed by the \texttt{Pre-Processor}.
However, continuous attitude adjustments with the reaction wheels,
orbital maneuvers and certain experiments involving the GRS or
the thrusters may cause the long term interruptions in the data. During
the time, the modeled data of the non-gravitational forces are needed
to fuse with the GRS measurements in order to fill the data-gaps.

As discussed in Sec. \ref{sec:Required-data-set}, the air-drag is
modeled as
\begin{equation}
\boldsymbol{a}_{drag}=-\frac{1}{2}\rho\frac{C_{D}A}{m_{SC}}\boldsymbol{v}_{rel}^{2}\frac{\boldsymbol{v}_{rel}}{\lvert\boldsymbol{v}_{rel}\rvert}\label{eq13}
\end{equation}
where $\rho$ is the air density, $C_{D}$ is the drag coefficient
($C_{D}=2.2$), $m_{SC}$ is the mass of the satellite, $\boldsymbol{v}_{rel}$
the velocity of the satellite relative to air, and $A$ is the windward
area of the satellite. The radiation pressure either from the Solar
radiation or the Earth albedo is modeled as
\begin{equation}
\boldsymbol{F}=\sum_{i}^{n}\boldsymbol{E}\frac{A_{i}\cos\theta}{c\,\,m_{SC}}\left[\left(1-\eta\right)\boldsymbol{\hat{E}}+2\eta\cos\theta\boldsymbol{\hat{n}}\right]\label{eq14}
\end{equation}
where $\boldsymbol{E}$ is the radiation flux, $A_{i}$ and $\boldsymbol{\hat{n}}$
denote the area and the normal direction of the S/C surface $i$,
$\eta$ the fraction of the incoming radiation that is reflected by
the surface $i$, and $\theta$ the angle between $\boldsymbol{E}$
and $\boldsymbol{\hat{n}}$. The key parameters in the above equations,
the drag coefficient and the reflection coefficient, depend on the
physical properties of the S/C surfaces and the real-time status of
the surrounding environments, and may also change with time. Therefore,
the modeled data is compared with the GRS measurements to adjust the
parameters. In Fig. \ref{fig:SimACC}, we calibrate the simulated
non-gravitational forces by the measured data from 01-11-2019. The
magnitude of the simulated atmospheric drag is smaller than that of
the Solar radiation pressure because of the 600 km altitude. In addition,
one could also see that the averaged Solar radiation pressure along
the pitch-axis (y-axis) of the satellite frame is the largest, since $\hat{\boldsymbol{y}}$
is the orbital normal direction oriented to the Sun.

\begin{figure}[ht]
\centering \includegraphics[width=0.58\textwidth]{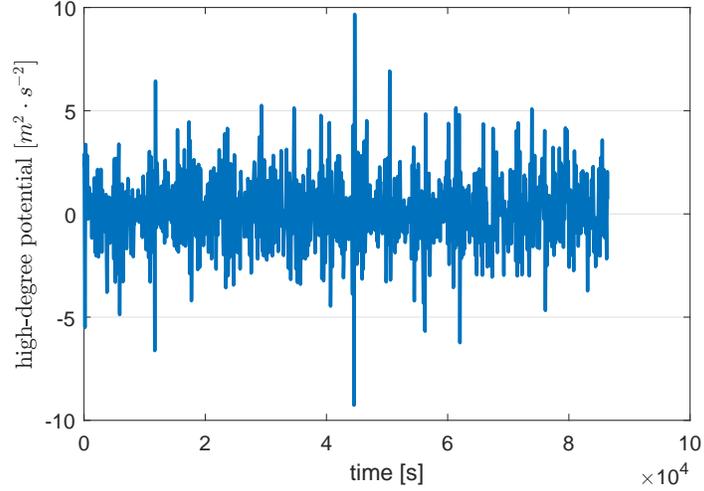} \caption{Time series of the high-degree geopotentials derived from EGM2008
for the spherical harmonic degree range 21 to 100 on 01-11-2019.(The greatest fluctuation around 44000 sec arises from transect across the Himalaya.) \label{fig:high-gravity}}
\end{figure}

Last but not the least, the high-degree geopotentials of degree/order
from 21 to 100 along the Taiji-1's orbit are derived from EGM2008.
See Fig. \ref{fig:high-gravity} for the illustration.

\begin{figure}[ht]
\centering \includegraphics[width=0.56\textwidth]{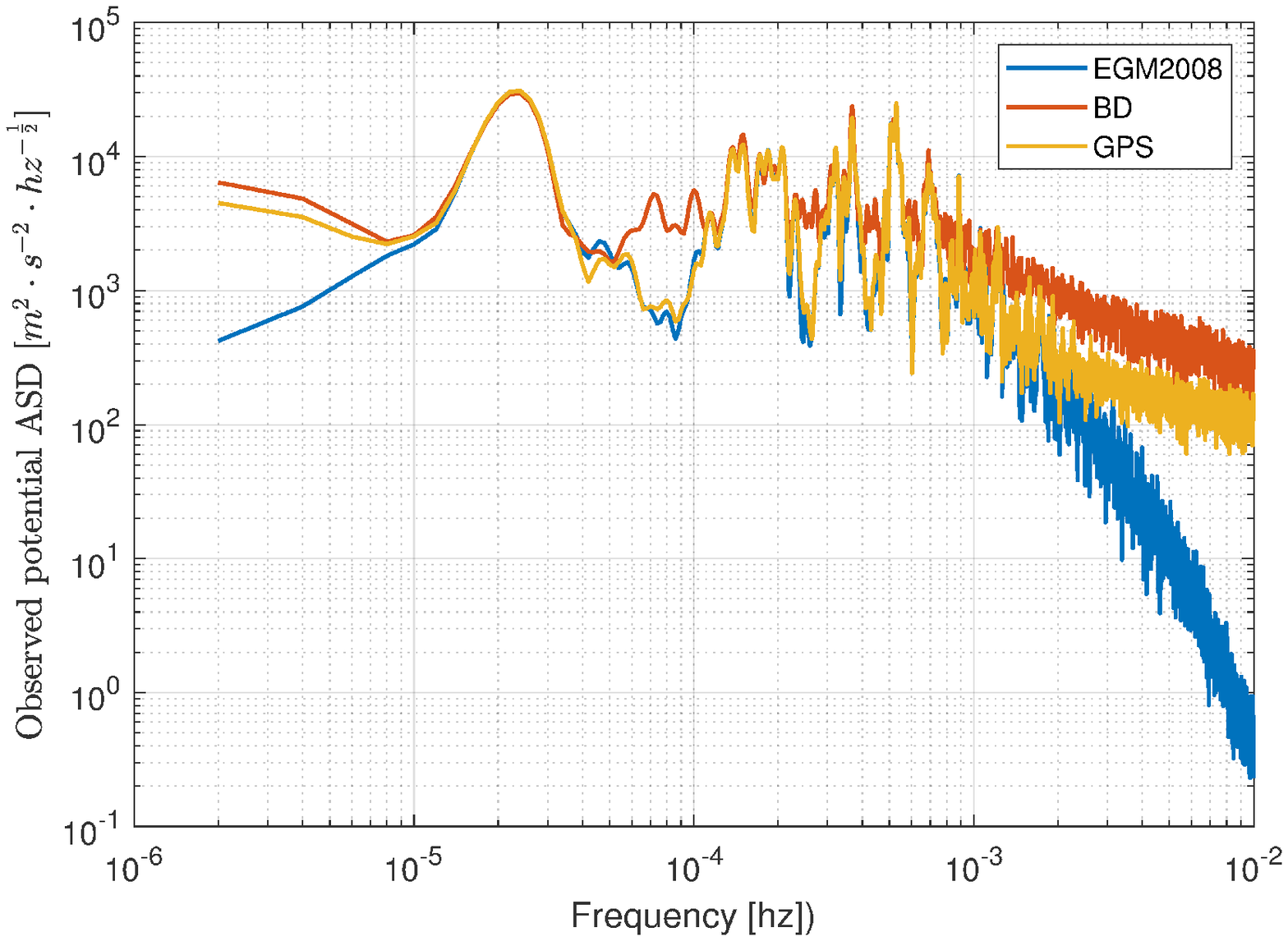} \caption{Comparisons of the ASD curves of the observed geopotentials and those based on the EGM2008 model along the Taiji-1's orbit. \label{fig:ASDgeopotential}}
\end{figure}

Given all the processed measurement data and the modeled ancillary
data so far, the observed low-degree Earth geopotentials along the
orbit is solved by the \texttt{SST-Processor} based on the observation
equation \ref{eq3}. In Fig. \ref{fig:ASDgeopotential}, the ASDs
of the geopotentials along the orbit derived from both the
BD and GPS measurements are compared with the simulated one based on the EGM2008 model.
It is found that the geopotentials from the GPS agree quite well with the EGM2008 model for $f<$3.5 mHz.
While, for frequency greater than 3.5 mHz, both of the BD and GPS results deviate significantly from the EGM2008 model. For Taiji-1's orbit, the frequency of 3.5 mHz corresponds to about the 20th-degree spherical harmonics.
This in fact indicates the precision limits of our gravity recovery method given the selected data sets.
Therefore, the low-pass filter at 5 mHz is imposed on the POD data, and the signals from the high-degree geopotentials of degree/order from 21 to 100 are modeled and subtracted from the observations in advance to determine the low-degree gravity field up to degree/order 20.

Finally, in the \texttt{Solver} sub-modular, the normal equations
based on Eq. \ref{eq1} are constructed in the form

\begin{equation}
\boldsymbol{Ax}=\boldsymbol{b}\label{eq10}
\end{equation}
where $\boldsymbol{b}$ is the observation vector, $\boldsymbol{x}$ is consisted of the unknown geopotential coefficients,
and $\boldsymbol{A}$ is the design matrix. The constrained gravity model is obtained in terms of
the least-square solutions of the above normal equations with
the first order Tikhonov regularization
\begin{equation}
\boldsymbol{x}=\left(\boldsymbol{A}^{\top}\boldsymbol{A} + \alpha \boldsymbol{K}\right)^{-1}\boldsymbol{A}^{\top}\boldsymbol{b}.\label{eq11}
\end{equation}
The elements $K_{ij}$ of the first order Tikhonov regularization matrix \textit{$\textbf{K}$} read
\begin{equation}
K_{ij}=\delta_{ij}n(i)\left(n(i)+1\right),\label{eq19}
\end{equation}
where $n(i)$ is the degree $n$ of the element in the row $i$.

\begin{figure}[ht]
\centering \includegraphics[width=0.46\textwidth]{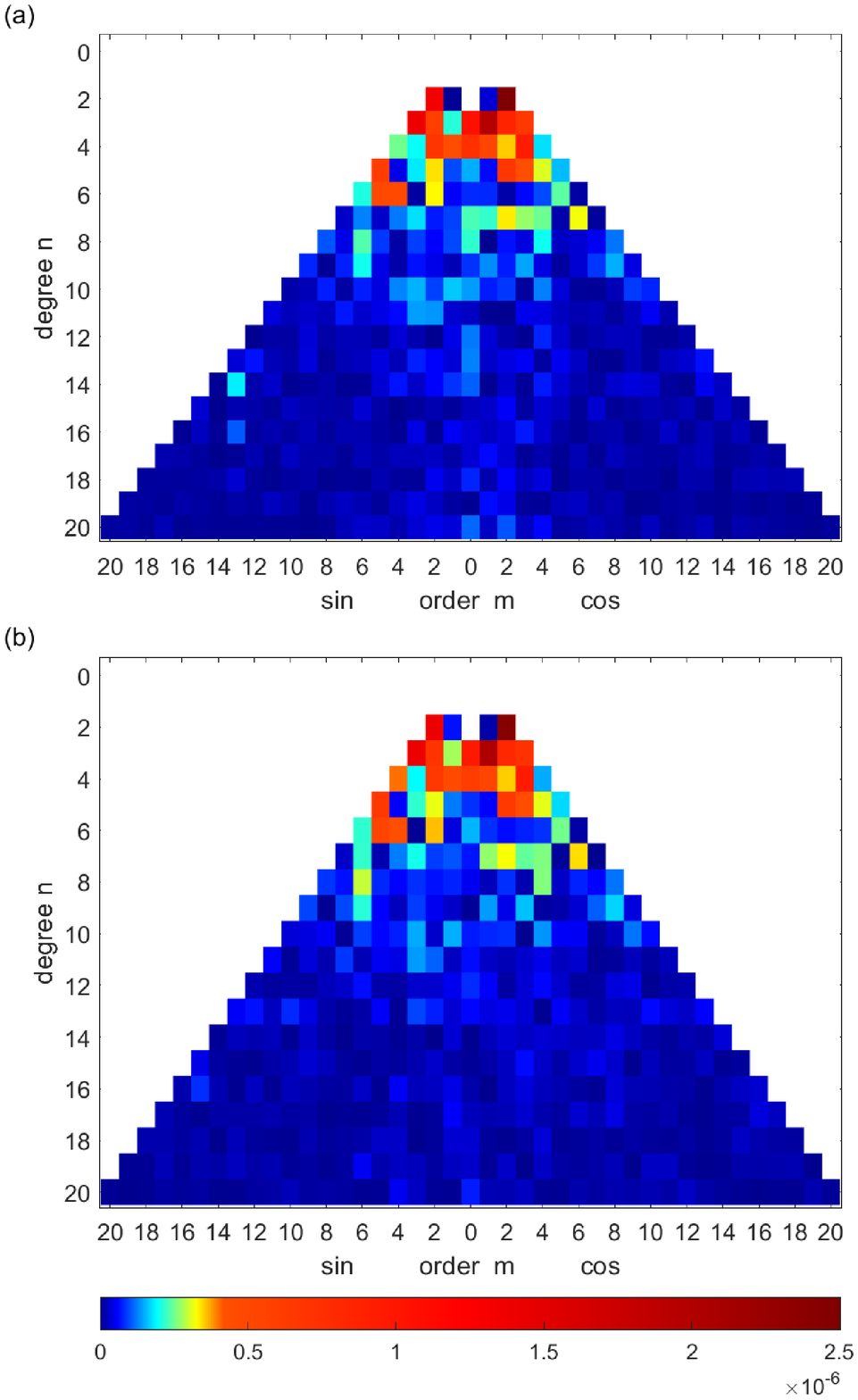}
\caption{Geopotential coefficients of the Taiji-1's gravity filed model \texttt{TJGM-r1911}
with the $C_{20}$ term removed: (a)BD; (b)GPS.\label{fig:ceo}}
\end{figure}

\begin{figure}[ht]
\centering \includegraphics[width=0.46\textwidth]{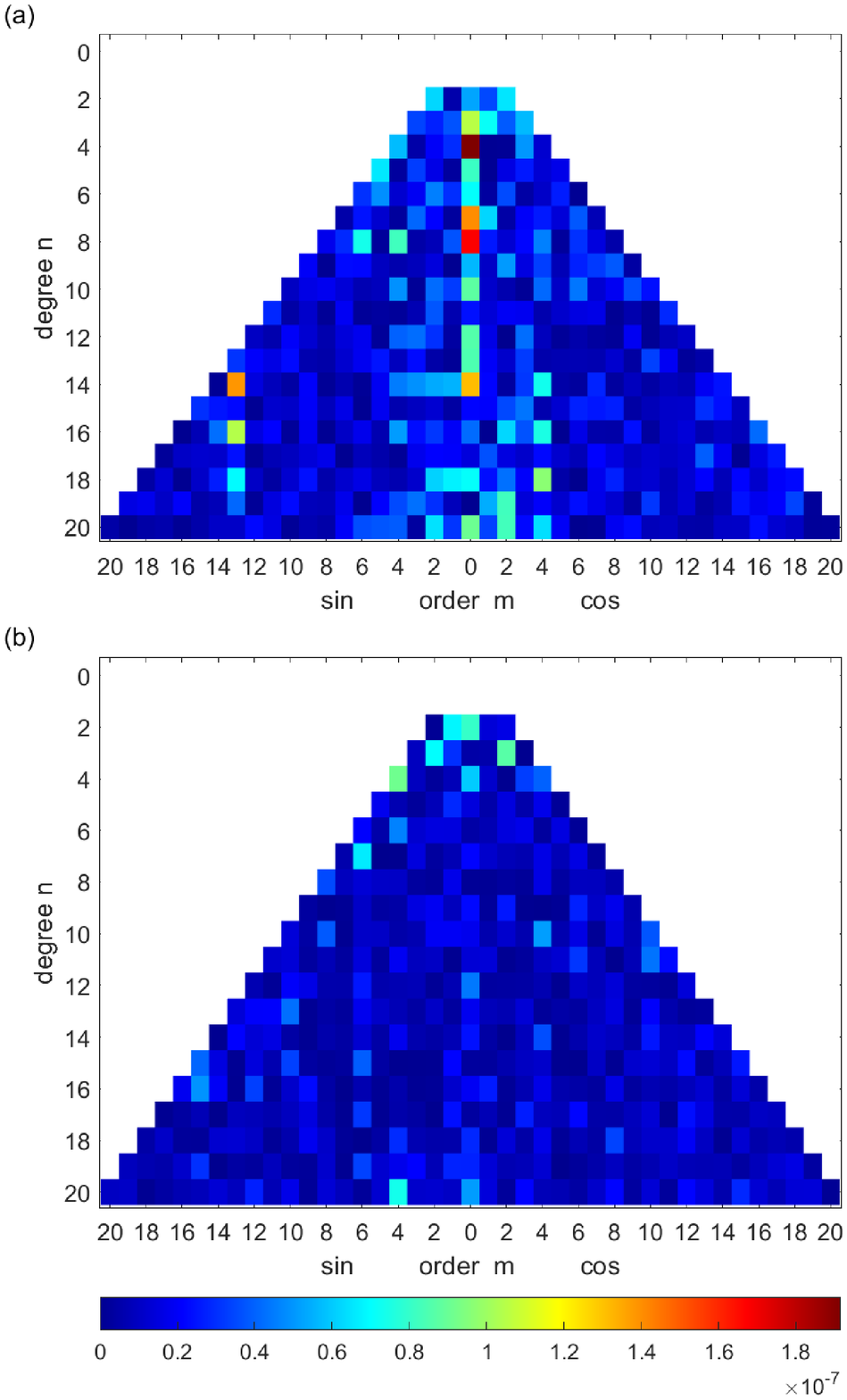}
\caption{Differences between the Taiji-1 gravity model \texttt{TJGM-r1911}
and EGM2008: (a)BD; (b)GPS. \label{fig:coediff}}
\end{figure}

\section{Results \label{sec:Results}}
We have obtained the first monthly global gravity model from Taiji-1's
observations based on the measurements in November 2019. The geopotential
data product, denoted as \texttt{TJGM-r1911}, is now archived by the
Taiji-1 data processing center of CAS at Beijing, and it will be released
soon together with the new monthly data products derived from the
much better observations during Taiji-1's extended free-falling phase
in this year. These data products contain the geopotentials
coefficients of spherical harmonics up to certain degree $\{C_{nm},\ S_{nm}\}$
and the corresponding variances $\{\Delta C_{mn},\ \Delta S_{nm}\}$.
\begin{figure}[ht]
\centering \includegraphics[width=0.58\textwidth]{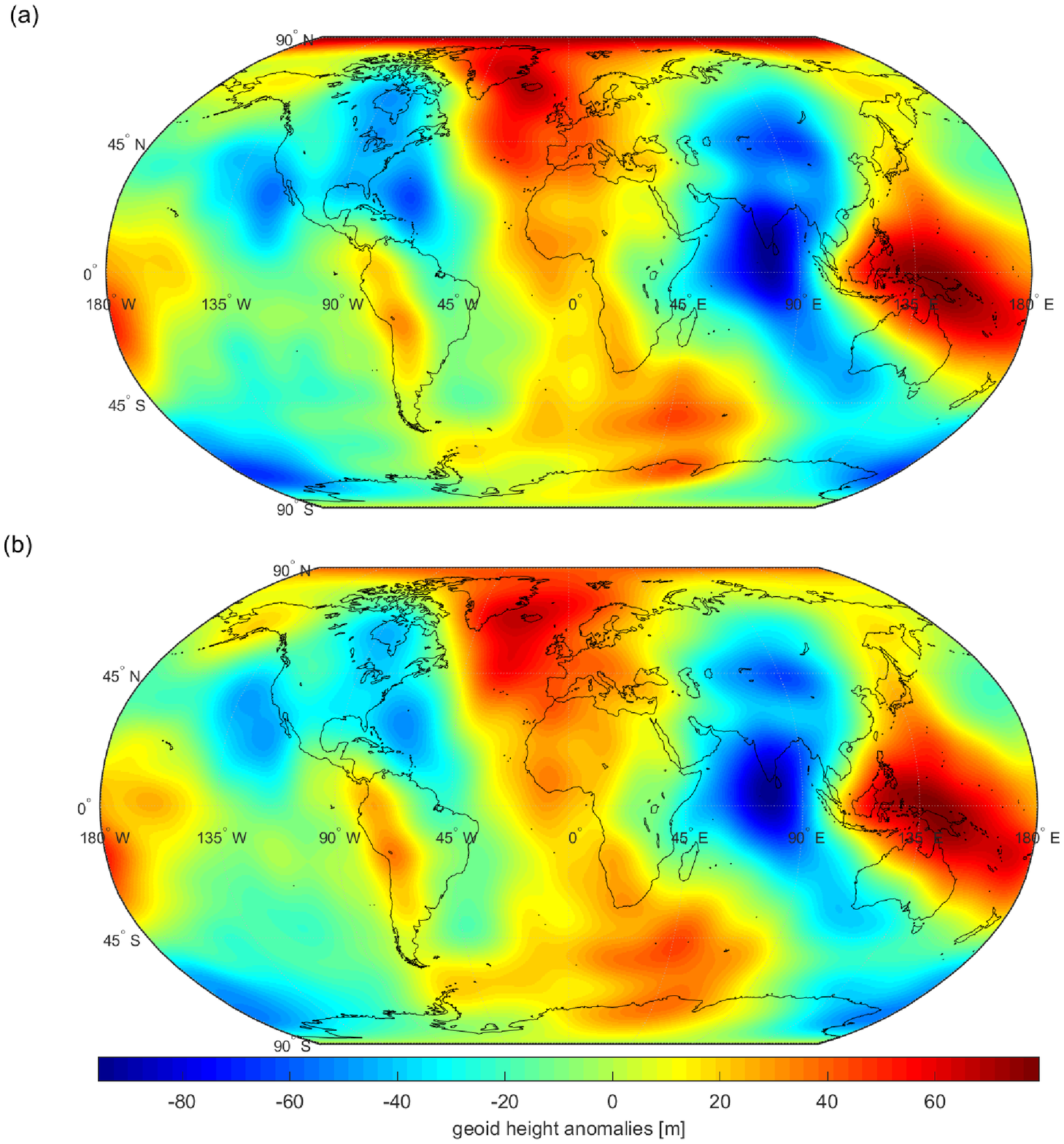}
\caption{Geoid height anomalies of the Taiji-1 gravity model \texttt{TJGM-r1911}: (a)BD;
(b)GPS. \label{fig:global}}
\end{figure}
\begin{figure}[ht]
\centering \includegraphics[width=0.58\textwidth]{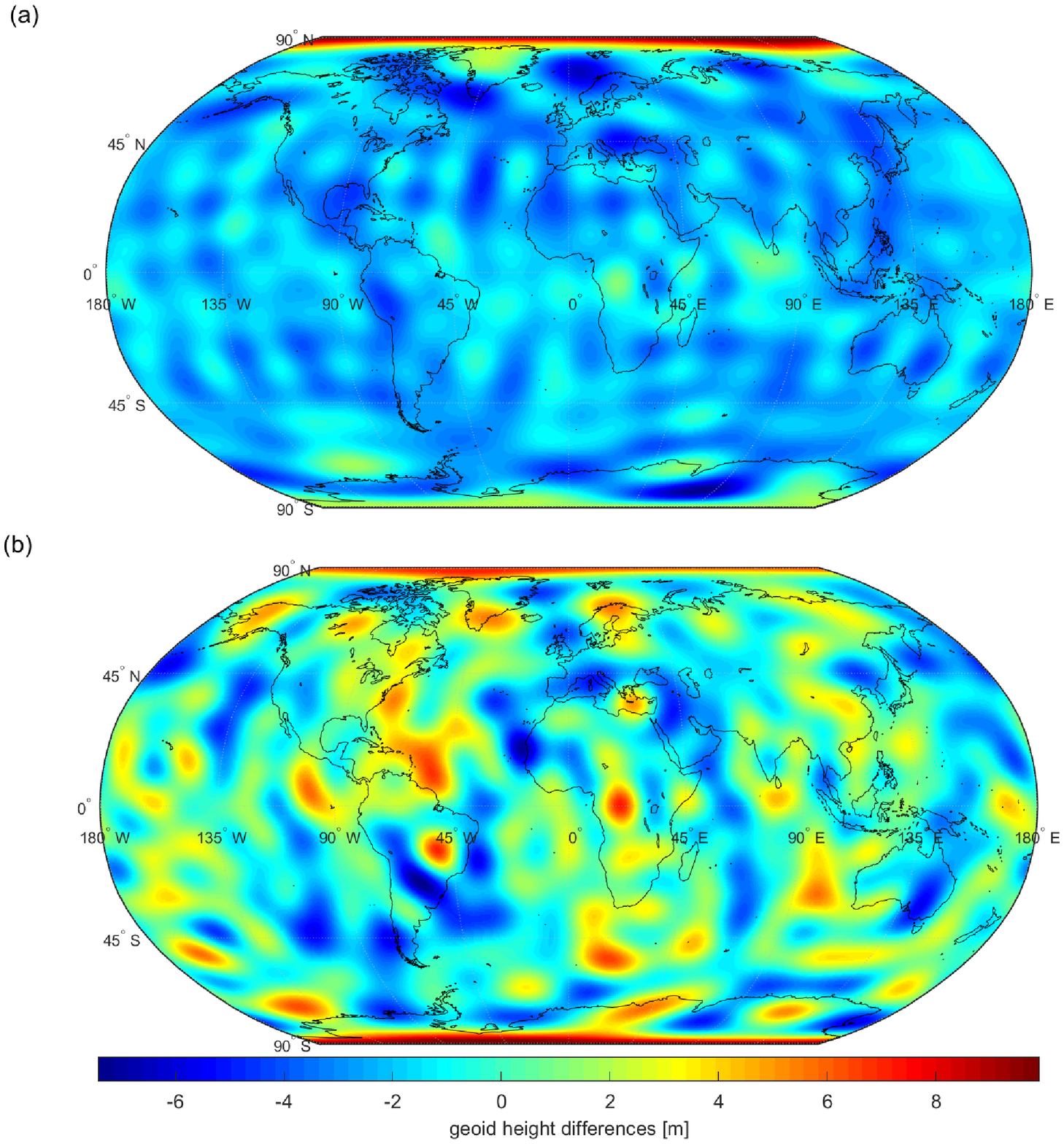}
\caption{Geoid height differences between the Taiji-1 gravity model \texttt{TJGM-r1911}
and EGM2008: (a)BD; (b)GPS.\label{fig:globaldiff}}
\end{figure}

\begin{figure}[ht]
\centering
\includegraphics[width=0.58\textwidth]{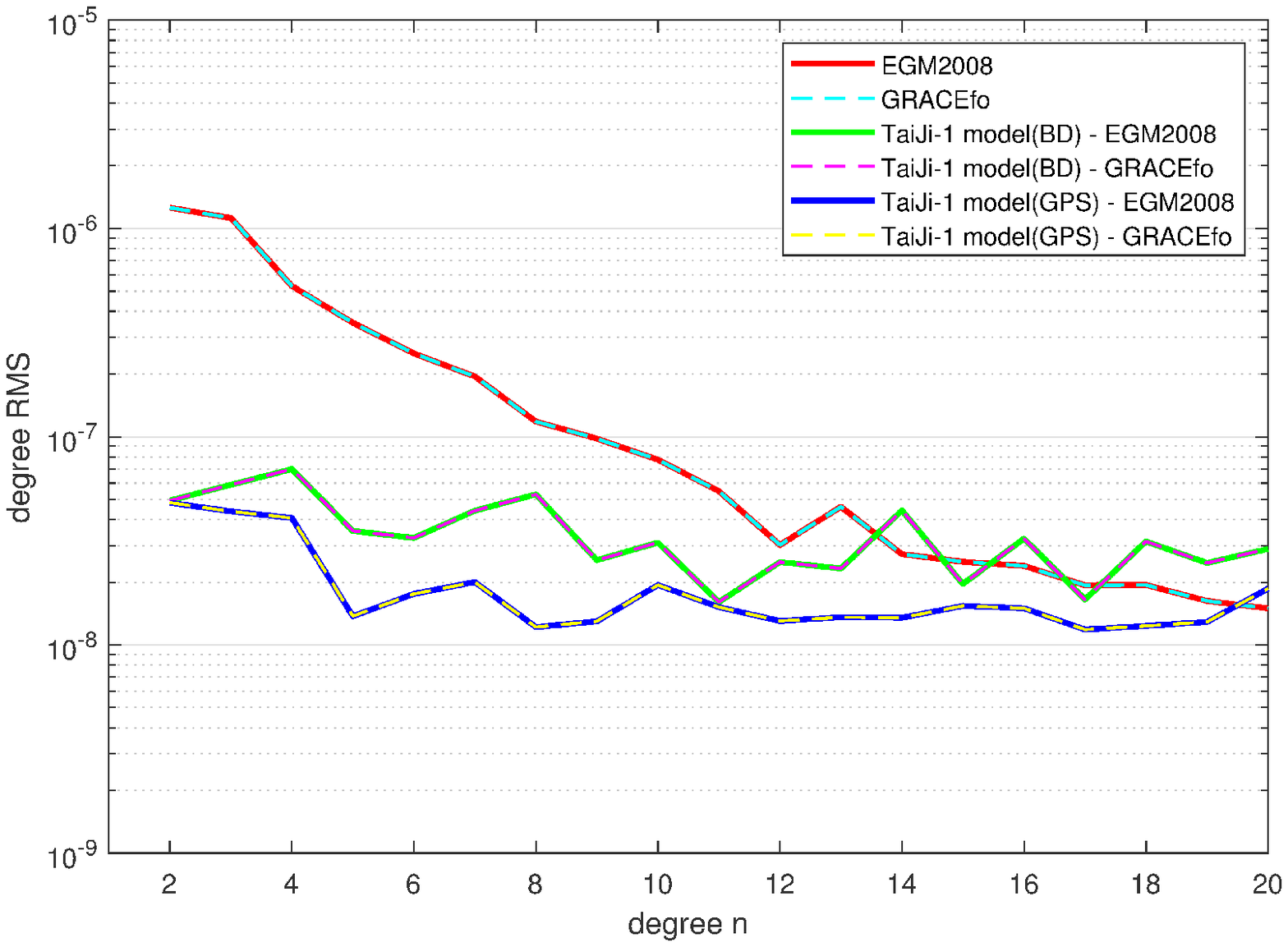}
\caption{Degree RMS of the Taiji-1 gravity model error. \label{fig:RMS}}
\end{figure}

As mentioned previously, mainly because of the long term interruptions
in the observation data, as well as the rather high orbit attitude
of Taiji-1, our first product \texttt{TJGM-r1911} is truncated at
the 20th degree. To make the cross-check and comparisons, \texttt{TJGM-r1911
}also contains two independent sets of the geopotential data obtained
from the GPS's and BD's observations respectively. Please see Fig.
\ref{fig:ceo} and Fig. \ref{fig:global} for the comparisons. The
differences from the reference models can be seen clearly in the plots in Fig. \ref{fig:coediff}, Fig. \ref{fig:globaldiff} and Fig. \ref{fig:RMS}. The gravity model
from BD observations shows larger deviations from the reference models
than that from the GPS observations.
Also, in Fig. \ref{fig:RMS}, the errors in terms of degree Root Mean Square
is estimated from the prior reference model $\{C_{nm}^{P},\ S_{nm}^{P}\}$
\begin{equation}
RMS_{n}=\sqrt{\frac{\sum_{m=0}^{n}\left(C_{nm}-C_{nm}^{P}\right)^{2}+\sum_{m=1}^{n}\left(S_{nm}-S_{nm}^{P}\right)^{2}}{2n+1}},\label{eq9}
\end{equation}
the deviations of the\texttt{ TJGM-r1911} model from the reference models
turns out to be larger than those from similar hl-SST missions. Compared
with the gravity models from the CHAMP mission \citep{gerlach_champ_2003,gerlach_champonly_2003,weigelt_timevariable_2013,reigber_gravity_2005,mayer-grr_itg-champ01_2005,han_efficient_2002,badura_derivation_2006,naeimi_global_2017},
the degree RMS of our \texttt{TJGM-r1911} model is about one order
of magnitude worse than that of the CHAMP models, depending on the
used methods and ancillary data. The poor degree RMS comes from two
factors. First, the Taiji-1's orbit has a large inclination angle,
and the altitude ($\sim600$ km) is higher than all the gravity recovery
satellites ever launched. Considering the key objectives of the Taiji-1 mission, this higher altitude was chosen to reduce the disturbances to the technology demonstration experiments from the space environment, especially the air drags.
This inclination causes signals loss in the
high latitude areas, and the higher altitude causes the magnitudes
of the geopotentials at the orbit to decrease. Second, as already
mentioned, the data is from the window period of the science operation
phase, even though the data set is carefully selected, there still
exist long interruptions and large disturbances, especially those
for the key GRS measurements of the non-gravitational forces.
The techniques such as data fusions with the calibrated simulations are
employed to improve the fitting accuracy. However, the problem of
the data quality and the discontinuity are believed to be the key
reasons to the increased errors. Till the preparation of this work, the data from October  2019 to May 2020 and September 2020 to May 2021 had been analyzed. The missing months from June  2020 to August 2020 are due to the encounters of the satellite with the Earth's shadow.
During these months, because of the power supply issue, the key payloads including the GRS system were turned off to reduce risks.
In the first few months until November 2019, instruments tests, calibrations and performance evaluations were still in progress, and in the following months much more experiments were scheduled. The precision of the geopotential product from October 2019 is comparable with that of \texttt{TJGM-r1911}, while the data from November 2019 is selected since it contains longer total operation time of the GRS system. For the subsequent months, the geopotential data qualities are rather worse.
While, all these data discontinuities and anomalies will be greatly improved in
the extended free-falling phase of Taiji-1 this year.

\section{Conclusions and outlook \label{sec:Conclusions}}
Gravity field is one of the fundamental physical
fields of a planet. For Earth, the global gravity field contains
vast amount of valuable information about the mass distributions and
transfers on the Earth surface, the global climate changes, the groundwater
storage, the Earth internal activities and so on. Great efforts have
been paid in the satellite gravity, and after the CHAMP, GRACE, GOCE,
and GRACE FO missions, the next generation gravity missions with advanced
technologies including relativistic measurements are under investigations.

In this work, we carefully select the data of one month from Taiji-1's
observations, and develop the data fusion techniques to resolve the
problem of the long term interruptions and the disturbances in the
measurements caused by the scheduled technology demonstration experiments.
The first global gravity model \texttt{TJGM-r1911} independently derived
from China's own satellite mission is then successfully produced from
the Taiji-1's observations. The existed discrepancies between the
first global gravity model \texttt{TJGM-r1911} and the gravity models
from CHAMP or other hl-SST missions are mainly caused by the data
discontinuity problems. The capability of the Taiji-1 satellite
serving as a fully-functional hl-SST satellite gravity mission for
China has been demonstrated and confirmed. The extended free-falling
phase of Taiji-1 with minimal disruptions and disturbances has been
approved to be started this year. With the techniques developed in
this work, Taiji-1 could serve as a satellite gravity mission
for China, which could provide us independently both the static mappings
of global gravity field and monthly averaged measurements of the time-variable
gravity field.

\begin{acknowledgments}
This work is supported by the National Key Research and Development
Program of China No. 2020YFC2200601 and No. 2020YFC2200104, the Strategic
Priority Research Program of the Chinese Academy of Sciences Grant
No. XDA15020700, and the Youth Fund Project of National Natural Science
Foundation of China No. 11905017.
\end{acknowledgments}


\section*{Declarations}

\textbf{Conflict of Interest} The authors have no competing interests to declare that are relevant to the content of this article.

\bibliographystyle{apsrev}
\bibliography{TJGrav}% common bib file

%\nolinenumbers
\end{document}